\documentclass{article} 
\usepackage{iclr2015,times}
\usepackage{hyperref}
\usepackage{url}
\usepackage{graphicx}
\usepackage{amsfonts}

\title{Weakly supervised multi-embeddings\\
learning of acoustic models}

\author{
Gabriel Synnaeve \& Emmanuel Dupoux \\ 
LSCP ENS/EHESS/CNRS\\
29 rue d'Ulm\\
75005, Paris, France \\
\texttt{gabriel.synnaeve@gmail.com}\\
\texttt{emmanuel.dupoux@gmail.com}
}

\begin{document}

\maketitle

\begin{abstract}
We trained a Siamese network with multi-task same/different information on a speech dataset, and found that it was possible to share a network for both tasks without a loss in performance. The first task was to discriminate between two same or different words, and the second was to discriminate between two same or different talkers.
\end{abstract}

\section{Introduction}

Theoretically, algorithms performing  unsupervised or weakly supervised discovery of linguistic structure represent plausible models of language acquisition in the human infant \citep{vallabha2007unsupervised}. Practically, they can be put to use for low resource languages \citep{park_unsupervised_2008}. 

Building on the fact that infant can recognize some words \citep{bergelson2012} and discriminate between speakers \citep{johnson2011infant} before they have constructed adult-like phoneme representations, we propose to test a neural network architecture where word and talker identity are used as side information to help learning an acoustic model (phone embedding). Previous work has shown that same-different side information can be used for metric learning \citep{xing2003}, and \cite{synnaeve2014SLT} demonstrated that it can be used with Deep Neural Network (DNN) architecture for learning phone embeddings. Here, we extend this work using multi-task (word and talker identity) side information.
As this paper is a feasibility study, we used gold same-different labels, and leave it to further work to derive them in an unsupervised fashion using spoken term discovery \citep{spoken_terms_discovery} and talker diarization \citep{anguera2012speaker}.
\section{Model}

We used the architure of a Siamese network \citep{bromley1993signature}, as shown in Fig.~\ref{fig:modeltrain}. It is a duplicated feedforward neural network taking two inputs in parallel. Each of the inputs consists in 11 stacked frames of 40 coefficients log-compressed Mel-filterbanks. Each network contains 3 hidden layers of 500 units with sigmoid activations, and two output embeddings each of 100 dimensions. One of the embeddings is the one in which we compute the similarity between the two inputs according to the same/different ``word type'' indication, while the other looks at the same/different ``speaker'' indication. More formally:
$$x_A\ \mathrm{and}\ x_B \in \mathbb{R}^{11\times40}\ ;\ y_{A,W},\ y_{A,S},\ y_{B,W}\ \mathrm{and}\ y_{B,S} \in \mathbb{R}^{100}$$
The loss function that we use (for two inputs $x_A$ and $x_B$) is a simple sum of the \textsc{coscos$^2$} losses in each of the embeddings (see \cite{synnaeve2014SLT} for a comparison with other loss functions):
\begin{eqnarray*}
\mathcal{L}(A,B) = \mathcal{L}_{W}(A,B) + \mathcal{L}_{S}(A,B)
\end{eqnarray*}
with $W \in \{0,1\}$ (different or same word) and $S \in \{0,1\}$ (different or same speaker), both losses are similar (here for speakers):
\begin{eqnarray*}
\mathcal{L}_{S}(A,B) = S\times(1-\cos(y_{A,S}, y_{B,S})) + (1-S)\times(\cos^2(y_{A,S}, y_{B,S}))
\end{eqnarray*}

\section{Experiments}

\subsection{Dataset}
We used about 1/3rd (12 speakers) of the Buckeye corpus\footnote{\begin{scriptsize}\url{http://buckeyecorpus.osu.edu}\end{scriptsize}}, on which we performed a dynamic time-warping (DTW) alignment of pairs of same words, in the features space (filterbanks), exactly as in \citep{synnaeve2014SLT}. This consisted in doing 76407 pairs of long 
``same'' word (1057 types in total), said 1/4th of the time by the same speakers (we subsampled ``same word and different speakers'' pairs). During training, we also sample pairs of tokens coming from different words (often called negative sampling), with a ratio of pairs of same/different words of 1:1. This yields about 5M frames for pairs of same words and 4.3M frames for sampled pairs of different words. 

\begin{figure}[h]
\begin{center}
\includegraphics[width=0.42\columnwidth]{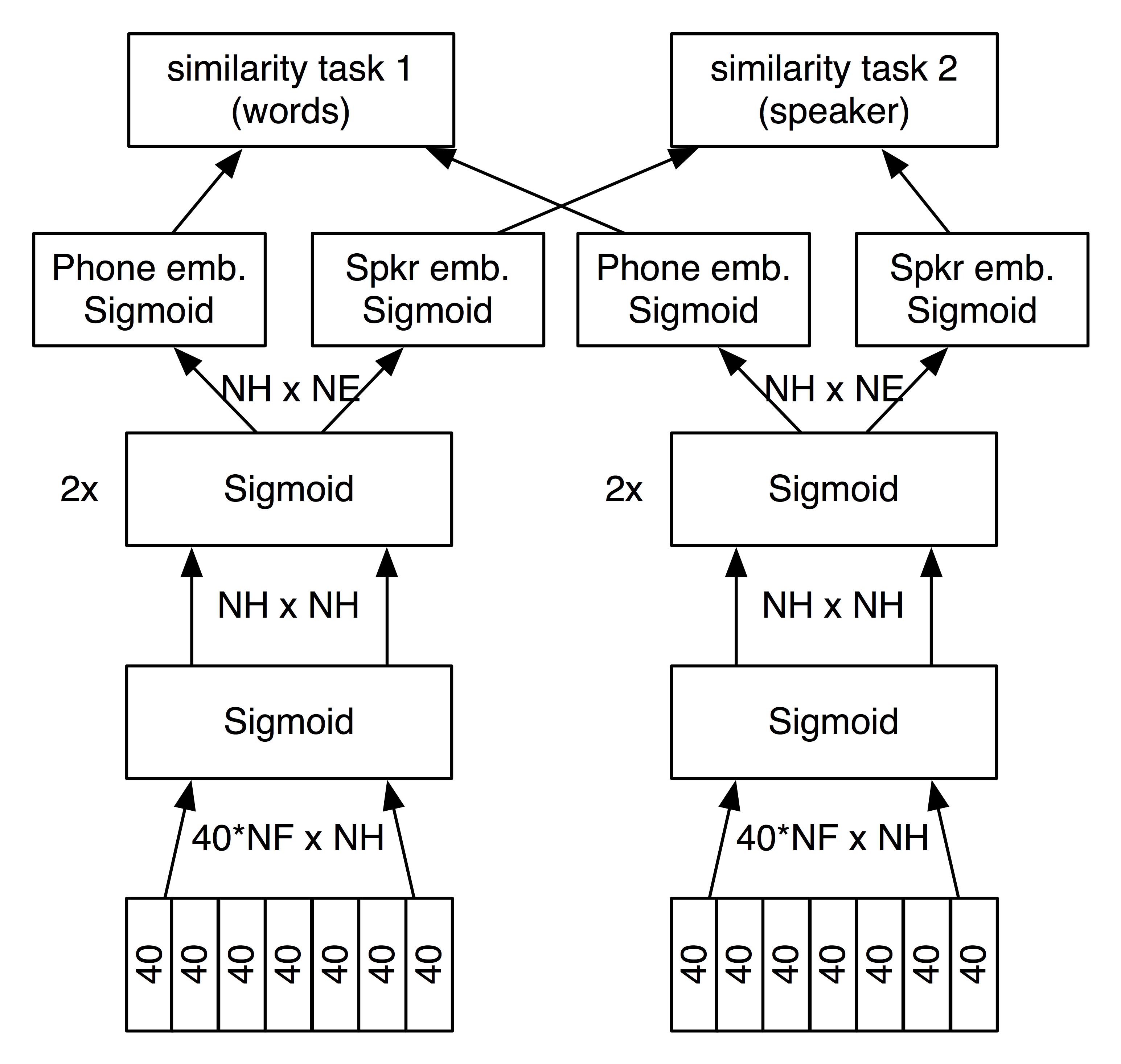}
\includegraphics[width=0.55\columnwidth]{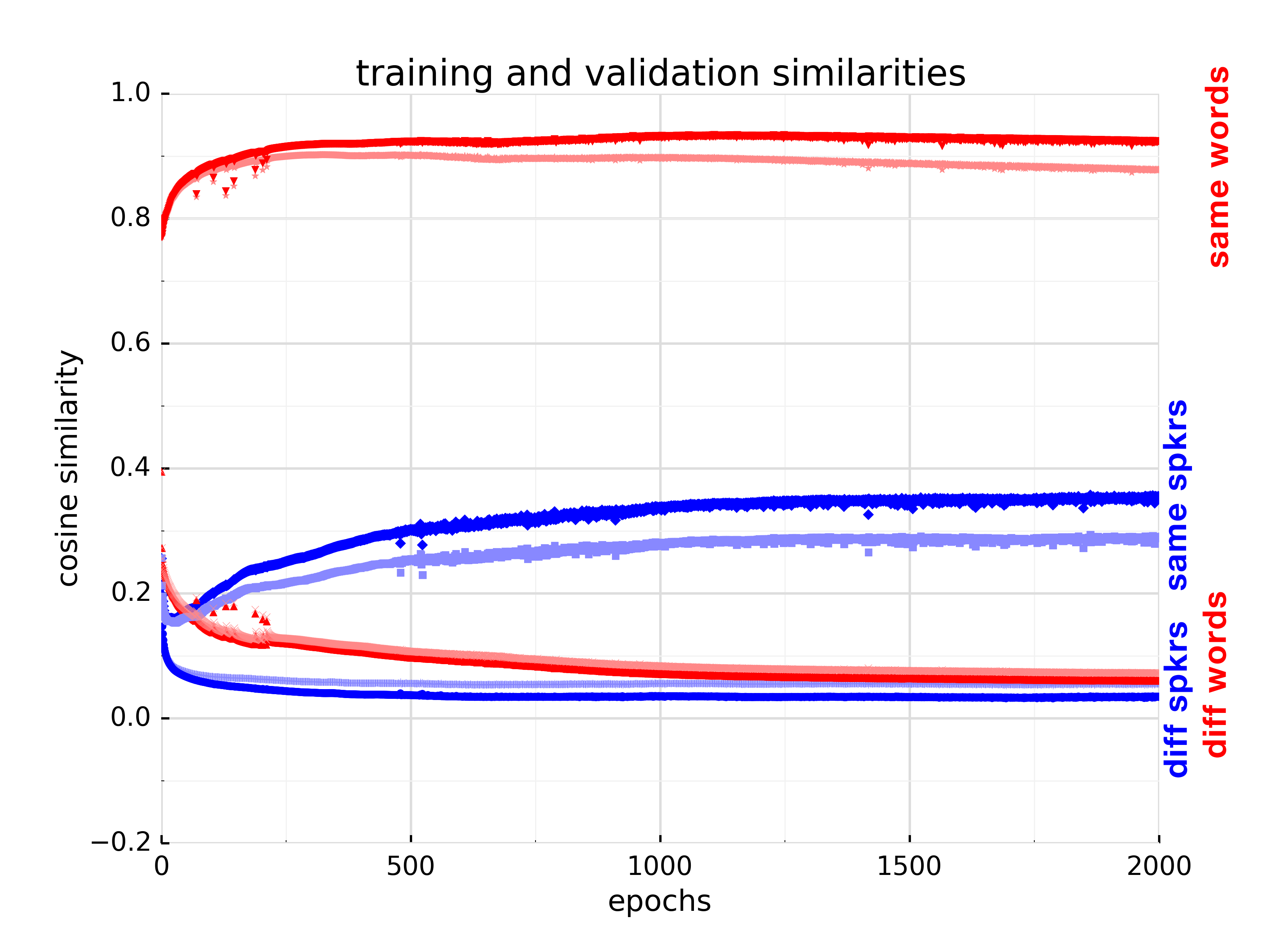}
\caption{Left: Architecture of our multi-embeddings learning Siamese network. We used NF=11, NH=500, and NE=100. Right: Evolution of cosine similarities for pairs of same/different words/speakers during training for the train set (saturated) and validation set (pastel).}
\end{center}
\label{fig:modeltrain}
\end{figure}

\subsection{Results}

We trained the model with Adadelta \citep{adadelta}, a variant of stochastic gradient descent with an adaptive learning rate method correcting the magnitude of the updates using an accumulation of past gradients (on a sliding window) and a local approximation of the Hessian. We used $\rho=0.95$ (hyper-parameter on the momentum) and $\epsilon=10^{-6}$ (precision of the updates), we performed early stopping on a small (10\%) held-out development set. We compared three network setups. In the multi-task setup, we use the combined loss function incorporating both the word and the speaker losses. In the single-loss setup, we only use one of the losses (word or speaker), even though the topology of the network remains the same. This means that the weights of only one of the two embeddings is updated, the other remaining in their initial state, thereby implementing a random projection from the last hidden layer. As a control, we also trained a fully supervised DNN using dropout \citep{srivastava2014dropout}. 
It has 11 stacked filterbank frames as input, 4 hidden layers of 2400 units, and 46 phones as outputs of the logistic regression (with a 37.9\% classification accuracy).

Figure \ref{fig:modeltrain} shows the evolution of the cosine similarities for the different conditions, and shows that the training of the speaker task takes more time than the training of the phoneme task, even though the cosine similarities start off less favorably for the former than the latter. In both cases, the difference between the train and the dev sets shows that the network is not really overfitting. 

Unsupervised systems do not necessarily discover phoneme-like units. Therefore, evaluating them with a phone error rate may not be appropriate. Similarly, using them as front end for a word recognizer may not be straightforward using standard HMMs. Here, we follow the lead of \cite{carlin2011rapid} and \cite{schatz2013} who propose to use instead a \emph{discrimination task}, which makes no assumption about the shape of the coded categories (phone-like, Gaussian, linearly separable, etc.) and does not depend on the training of a classifier or a language model. The ABX discrimination task consists in computing two pairwise distances, between the token pair X and A, and between X and B and deciding which of them is larger. When A and B are tokens of different linguistic categories, and X belongs to the category of either A or B, this metric measures the \emph{degree of separation} of the two categories A and B in the embedding. Here, we use as categories, minimal pairs of triphones of the shape /a/-/t/-/i/ vs /a/-/p/-/i/, where the left and right context phones are kept identical, and the center phone varies. As distance metric, we use the cosine distance along the DTW path. We setup two tasks, on which we will test our two embeddings:  
\begin{itemize}
\item \textit{phone.talker} is a phoneme discrimination task across speakers. For instance, A=/a/-/p/-/i/, B=/a/-/t/-/i/, both being said by the same speaker, and X is phonetically identical to A or B, but is uttered by a different speaker.
\item \textit{talker.phone} is a speaker discrimination task, across phonemes. For instance, both A and B share the same triphone (eg., /a/-/p/-/i/) but are said by different speakers; X is uttered by either the speaker of A or the speaker of B, but has different phonemes (eg., /a/-/t/-/i/).
\end{itemize}

The two tasks are mirror image of one another regarding the discrimination of the phonemes or of the talkers. In both cases, the context (left and right) phonemes are kept identical. To run this task, we select the set of all eligible ABX triplets of triphones in the dataset, and compute the aggregate ABX score by averaging across context, phoneme and speaker pairs.

\begin{figure}[h]
\begin{center}
\includegraphics[width=0.72\columnwidth]{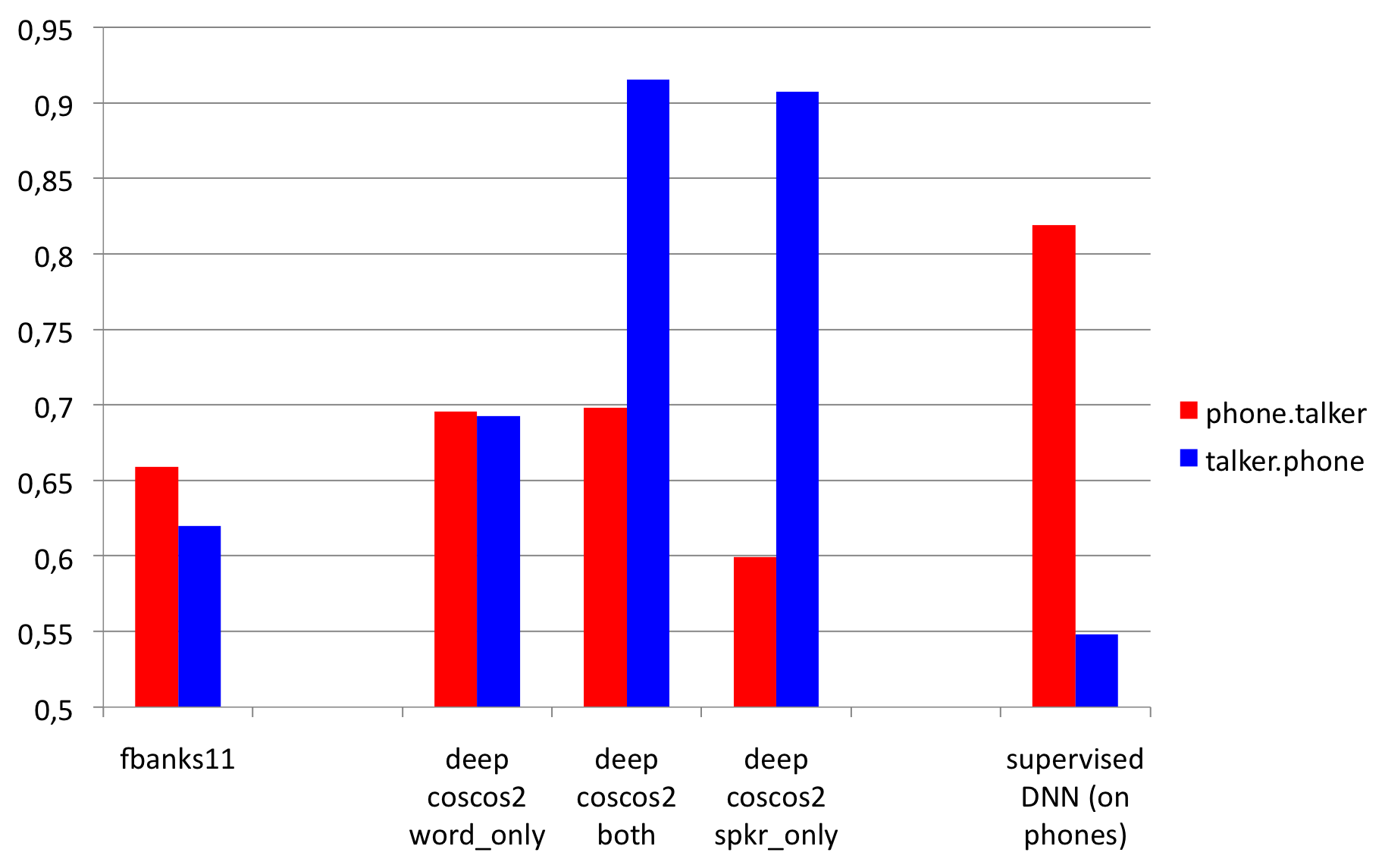}
\caption{ABX scores for speech features (11 frames of stacked filterbanks), for three Siamese networks: one trained with a same/different word loss function (``word\_only''), one with a multi-task loss (``both''), and one same/different speakers loss functions (``spkr\_only''). The three networks have the same topology and the ABX tasks (phone or speaker) are run, respectively, on the phone-based and the speaker-based embeddings. A control shows a supervised DNN trained on phones.}
\end{center}
\label{fig:ABXscores}
\end{figure}

\subsection{Discussion}
The ABX scores for the phoneme and speaker tasks are shown in Fig.~\ref{fig:ABXscores}. Globally, speaker discrimination seems easier to optimize than phoneme discrimination (even though it starts the other way around when evaluated from the filterbanks). This is probably due to the small number of speaker classes (N=12) compared to the number of phoneme classes (N=48). In addition, the multi-task network gives the best results across the two tasks, compared to single-task networks. Therefore, learning to do two tasks at once using the same network does not incur a decrease in performance, but on the contrary is slightly beneficiary (especially for the talker task). Interestingly the single-task networks behave asymmetrically with respect to the untrained task. Indeed, the performance on phone discrimination is worse for the network that was trained only on the speaker loss, compared to the filterbank performance. This makes sense: if you are trained to ignore phoneme identity, phoneme encoding should be progressively removed from the hidden layers of the network. But vice-versa the performance on speaker discrimination is \emph{better} for the phoneme-loss network compared to the filterbank base performance. This means that in order to determine speaker identity, it is actually useful for the network to code some information about the phonemes. This last result meshes well with the fact that speaker identification depends not so much on raw acoustic features, but on small deformations relative to a background pronunciation distribution (as encoded in \textit{i-vectors}, \cite{dehak2011front}). Specialization on the task is even more extreme for the fully supervised DNN trained on phone labels: it gives us a higher bound on the \textit{phone accross talkers} task (81.9\% correct), and shows degraded \textit{talker accross phones} score (54.8\% correct) compared to the filterbank. 

\begin{figure}[h]
\begin{center}
\includegraphics[width=0.78\columnwidth]{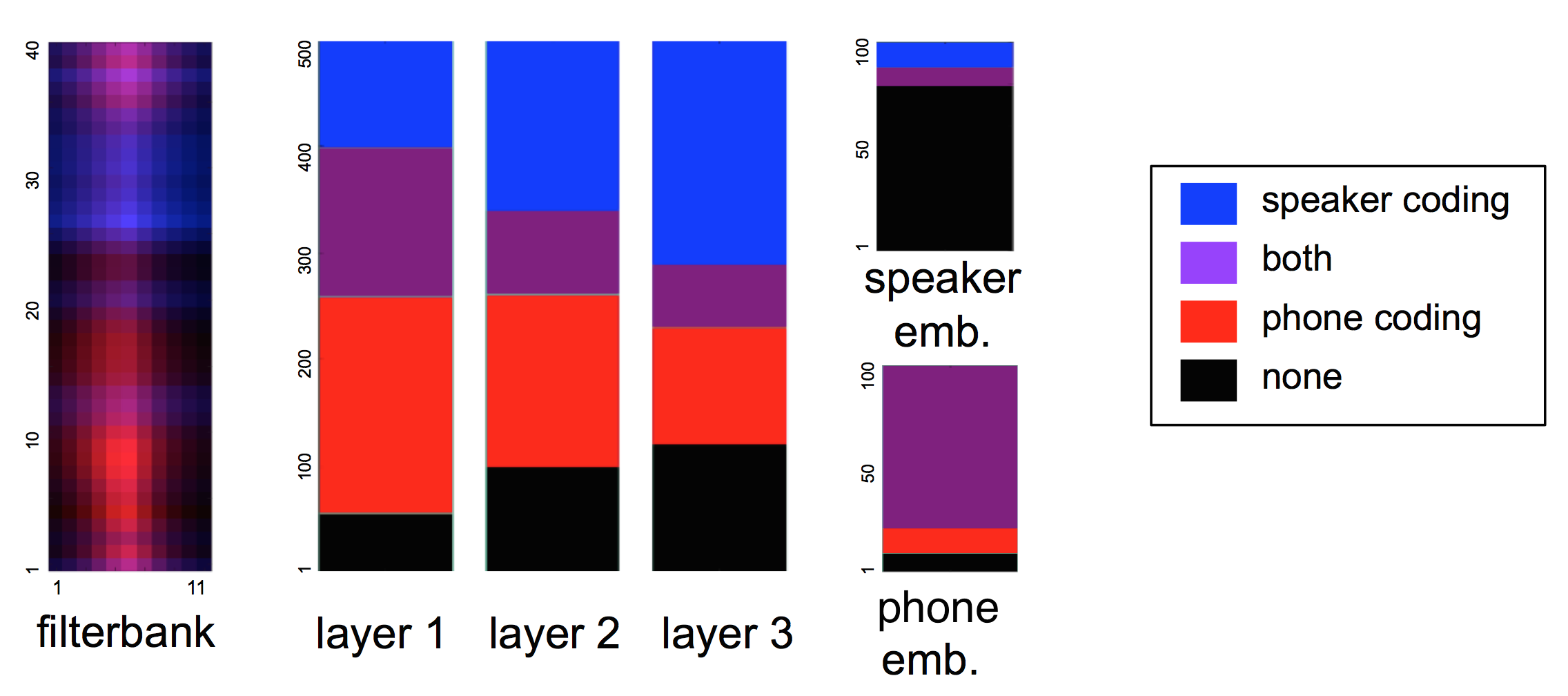}
\caption{Coding specificity of the input, hidden and embedding layers of the AB net, computed using the ratio of between- to within-category variance (F-test). Left: 11 stacked filterbanks coding of speakers (shades of blue) and phones (shades of red). The x-axis represent the 11 stacked frames, the y-axis represents the 40 filterbanks coefficients. Right: Cumulative barplots representing the number of units in the layers coding specifically for  speakers (blue), phones (red), both phones and speakers (purple), or none (black). A unit is deemed code-specific if the between/within variance ratio for that category is more than the network-wide median.}
\end{center}
\label{fig:specificity}
\end{figure}
\vspace{-0.3cm}

In order to understand the nature of information encoding in a multi-layer network, it can be useful to inspect the hidden layers in details \citep{mohamed2012understanding}. Here, 
we inspected each hidden unit by computing the ratio of between-class to within-class variance in unit activation over the entire corpus (F-test). To compute the phoneme variance ratio, we took the variance of the activation value of the unit across all (between) the phone categories versus within each phone category. We did a similar computation for the speakers categories. Intuitively, a unit with a large phoneme variance ratio is strongly encoding phoneme information; a unit with a small ratio is not very sensitive to that information. Similarly for speaker information. If we split the ratio distribution using the median, this gives rise to a typology of 4 kinds of units according to whether they respond strongly or not to either phone or speaker information. In Figure 3, we can see three phenomena regarding the coding of units in the three hidden layers. First, phone-coding units are predominant in the first layers, and progressively, more and more speaker-coding units appear. Second, the number of doubly-coding units diminishes. Third, the sparsity of the code increases (ie., the number of units not coding anything). Inspection of the task-specific embeddings is interesting, as it reveals an almost pure (and very sparse) coding of speaker identity in the speaker embedding. This is consistent with the high performance of speaker discrimination in this layer. In contrast, inspection of the 
phone embedding reveals a much less sparse coding and a predominance of doubly-used units. This is consistent with the rather low performance of phoneme discrimination in this layer, and suggests that further layers or more (speaker variability in) training examples would be necessary to ``purge''  this layer from speaker-specific effects. 
Finally, inspection of the filterbanks (here coded in shade of red and blue) shows that most of the lower frequency filterbanks are sensitive to phone infrmation (relatively localized in time to the center frames, as we compute it on phonetic annotation), whereas the higher frequency filterbanks are sensitive to speaker information (relatively \textit{not} localized in time).

\section{Conclusion}
We have demonstrated that a Siamese network can perform both phoneme and speaker discrimination using only a moderate amount of side information (indication of same/different word or speaker for only $\approx$1000 word types and 12 speakers). Further work is needed to study the effect of the amount of information, and whether the obtained speaker embeddings could replace or complement \textit{i-vectors}. Finally, the phone embedding should be evaluated as a first step in a subsequent word recognizer or language model adapted for this kind of representation.

\subsubsection*{Acknowledgments}
This project is funded in part by the European Research Council (ERC-2011-AdG-295810 BOOTPHON), the Agence Nationale pour la Recherche (ANR-10-LABX-0087 IEC, ANR-10-IDEX-0001-02 PSL*), the Fondation de France, the Ecole de Neurosciences de Paris, and the Region Ile de France (DIM cerveau et pensée).

\bibliography{main}

\begin{thebibliography}{15}
\providecommand{\natexlab}[1]{#1}
\providecommand{\url}[1]{\texttt{#1}}
\expandafter\ifx\csname urlstyle\endcsname\relax
  \providecommand{\doi}[1]{doi: #1}\else
  \providecommand{\doi}{doi: \begingroup \urlstyle{rm}\Url}\fi

\bibitem[Anguera~Miro et~al.(2012)Anguera~Miro, Bozonnet, Evans, Fredouille,
  Friedland, and Vinyals]{anguera2012speaker}
Anguera~Miro, Xavier, Bozonnet, Simon, Evans, Nicholas, Fredouille, Corinne,
  Friedland, Gerald, and Vinyals, Oriol.
\newblock Speaker diarization: A review of recent research.
\newblock \emph{Audio, Speech, and Language Processing, IEEE Transactions on},
  20\penalty0 (2):\penalty0 356--370, 2012.

\bibitem[Bergelson \& Swingley(2012)Bergelson and Swingley]{bergelson2012}
Bergelson, Elika and Swingley, Daniel.
\newblock At 6--9 months, human infants know the meanings of many common nouns.
\newblock \emph{Proceedings of the National Academy of Sciences}, 109\penalty0
  (9):\penalty0 3253--3258, 2012.

\bibitem[Bromley et~al.(1993)Bromley, Bentz, Bottou, Guyon, LeCun, Moore,
  S{\"a}ckinger, and Shah]{bromley1993signature}
Bromley, Jane, Bentz, James~W, Bottou, L{\'e}on, Guyon, Isabelle, LeCun, Yann,
  Moore, Cliff, S{\"a}ckinger, Eduard, and Shah, Roopak.
\newblock Signature verification using a “siamese” time delay neural
  network.
\newblock \emph{Internat. Journ. of Pattern Recog. and Artific. Intell.},
  7\penalty0 (04):\penalty0 669--688, 1993.

\bibitem[Carlin et~al.(2011)Carlin, Thomas, Jansen, and
  Hermansky]{carlin2011rapid}
Carlin, Michael~A, Thomas, Samuel, Jansen, Aren, and Hermansky, Hynek.
\newblock Rapid evaluation of speech representations for spoken term discovery.
\newblock In \emph{INTERSPEECH}, pp.\  821--824, 2011.

\bibitem[Dehak et~al.(2011)Dehak, Kenny, Dehak, Dumouchel, and
  Ouellet]{dehak2011front}
Dehak, Najim, Kenny, Patrick, Dehak, R{\'e}da, Dumouchel, Pierre, and Ouellet,
  Pierre.
\newblock Front-end factor analysis for speaker verification.
\newblock \emph{Audio, Speech, and Language Processing, IEEE Transactions on},
  19\penalty0 (4):\penalty0 788--798, 2011.

\bibitem[Jansen et~al.(2010)Jansen, Church, and
  Hermansky]{spoken_terms_discovery}
Jansen, Aren, Church, Kenneth, and Hermansky, Hynek.
\newblock Towards spoken term discovery at scale with zero resources.
\newblock In \emph{INTERSPEECH}, pp.\  1676--1679, 2010.

\bibitem[Johnson et~al.(2011)Johnson, Westrek, Nazzi, and
  Cutler]{johnson2011infant}
Johnson, Elizabeth~K, Westrek, Ellen, Nazzi, Thierry, and Cutler, Anne.
\newblock Infant ability to tell voices apart rests on language experience.
\newblock \emph{Developmental Science}, 14\penalty0 (5):\penalty0 1002--1011,
  2011.

\bibitem[Mohamed et~al.(2012)Mohamed, Hinton, and
  Penn]{mohamed2012understanding}
Mohamed, Abdel-rahman, Hinton, Geoffrey, and Penn, Gerald.
\newblock Understanding how deep belief networks perform acoustic modelling.
\newblock In \emph{Acoustics, Speech and Signal Processing (ICASSP), 2012 IEEE
  International Conference on}, pp.\  4273--4276. IEEE, 2012.

\bibitem[Park \& Glass(2008)Park and Glass]{park_unsupervised_2008}
Park, Alex~S. and Glass, James~R.
\newblock Unsupervised pattern discovery in speech.
\newblock \emph{{IEEE} Transactions on Audio, Speech, and Language Processing},
  16\penalty0 (1):\penalty0 186--197, January 2008.
\newblock ISSN 1558-7916.
\newblock \doi{10.1109/TASL.2007.909282}.

\bibitem[Schatz et~al.(2013)Schatz, Peddinti, Bach, Jansen, Hynek, and
  Dupoux]{schatz2013}
Schatz, Thomas, Peddinti, Vijayaditya, Bach, Francis, Jansen, Aren, Hynek,
  Hermansky, and Dupoux, Emmanuel.
\newblock Evaluating speech features with the minimal-pair abx task: Analysis
  of the classical mfc/plp pipeline.
\newblock In \emph{{INTERSPEECH-2013}}, pp.\  1781--1785, 2013.

\bibitem[Srivastava et~al.(2014)Srivastava, Hinton, Krizhevsky, Sutskever, and
  Salakhutdinov]{srivastava2014dropout}
Srivastava, Nitish, Hinton, Geoffrey, Krizhevsky, Alex, Sutskever, Ilya, and
  Salakhutdinov, Ruslan.
\newblock Dropout: A simple way to prevent neural networks from overfitting.
\newblock \emph{The Journal of Machine Learning Research}, 15\penalty0
  (1):\penalty0 1929--1958, 2014.

\bibitem[Synnaeve et~al.(2014)Synnaeve, Schatz, and Dupoux]{synnaeve2014SLT}
Synnaeve, Gabriel, Schatz, Thomas, and Dupoux, Emmanuel.
\newblock Phonetics embedding learning with side information.
\newblock In \emph{IEEE SLT}, 2014.

\bibitem[Vallabha et~al.(2007)Vallabha, McClelland, Pons, Werker, and
  Amano]{vallabha2007unsupervised}
Vallabha, Gautam~K, McClelland, James~L, Pons, Ferran, Werker, Janet~F, and
  Amano, Shigeaki.
\newblock Unsupervised learning of vowel categories from infant-directed
  speech.
\newblock \emph{Proceedings of the National Academy of Sciences}, 104\penalty0
  (33):\penalty0 13273--13278, 2007.

\bibitem[Xing et~al.(2003)Xing, Ng, Jordan, and Russell]{xing2003}
Xing, Eric~P, Ng, Andrew~Y, Jordan, Michael~I, and Russell, Stuart.
\newblock Distance metric learning with application to clustering with
  side-information.
\newblock \emph{Advances in neural information processing systems}, pp.\
  521--528, 2003.

\bibitem[Zeiler(2012)]{adadelta}
Zeiler, Matthew~D.
\newblock Adadelta: An adaptive learning rate method.
\newblock \emph{arXiv preprint:1212.5701}, 2012.

\end{thebibliography}
\bibliographystyle{iclr2015}

\end{document}